\documentclass[useAMS,usenatbib]{mn2e}
\usepackage{epsf}
\usepackage{amssymb}
\usepackage[usenames]{color}

\newcommand {\bc}{\begin {center}}
\newcommand {\ec}{\end {center}}
\newcommand {\be}{\begin {equation}}
\newcommand {\ee}{\end {equation}}
\newcommand {\beq}{\begin {eqnarray}}
\newcommand {\eeq}{\end {eqnarray}}

\def\plotone#1{\centering \leavevmode
\epsfxsize=\columnwidth \epsfbox{#1}}

\def\disp {\displaystyle}
\def\e {{\rm e}}
\def\D {{\rm D}}

\def\i {{\rm i}}

\def\r {{\rm r}}

\title[Resonant scattering in galaxy clusters for gas motions]{Resonant
  scattering in galaxy clusters for anisotropic gas motions on various spatial scales}
\author[Zhuravleva et
  al.]{I.Zhuravleva$^{1}$\thanks{izhur@mpa-garching.mpg.de},
  E.Churazov$^{1,2}$, S.Sazonov$^{2,1}$, R.Sunyaev$^{1,2}$, K.Dolag$^{1}$\\ \\
$^{1}$Max Planck Institute for Astrophysics, Karl-Schwarzschild str. 1, Garching, 85741, Germany\\
$^{2}$Space Research Institute, Profsoyuznaya str. 84/32, Moscow,
  117997, Russia}

\begin{document}


\pagerange{\pageref{firstpage}--\pageref{lastpage}} \pubyear{2009}

\maketitle

\label{firstpage}

\begin{abstract}
The determination of characteristic amplitudes and anisotropy of hot gas
motions in galaxy clusters from observations of the brightest resonance lines is discussed. Gas motions
affect (i) the spectral line shape through the Doppler effect and (ii) the radial surface
brightness profiles in lines during resonant scattering. Radiative transfer calculations have been performed
by the Monte Carlo method in the FeXXV resonance line at 6.7 keV for
the Perseus cluster (Abell426). It was shown that (a) radial motions reduce the scattering efficiency much more
dramatically than purely tangential motions; (b) large-scale gas
motions weakly affect the scattering efficiency. The uncertainty in
measuring the characteristics of gas motions using
resonant scattering has been estimated for existing and future observations of clusters.  
\end{abstract}

\section{Introduction} 
\label{sec:intro}

The intergalactic gas in galaxy clusters is the
dominant (in mass) baryonic component and accounts
for about 15\% of the virial cluster mass.
Falling into the potential well of a cluster, the gas
heats up to a temperature  of 2--10 keV. The presence
of gas motions with velocities of hundreds and
thousands km s$^{-1}$ in galaxy clusters is indirectly
confirmed both by high-angular-resolution X-ray data, namely the
observations of substructure in surface brightness and temperature distributions (e.g. see the review by \citet{Mar07}),
and by numerical simulations, which show gas motions
on various spatial scales, both far from the
center and in the central regions of clusters (\citet{Nor99,Ino03,Dol05,Vaz09}).

The anisotropy, velocity amplitudes and scales of
the motions along the line-of-sight can be determined
from the shift and broadening of spectral lines
(e.g. see \cite{Ino03}). However,
measurements with a sufficient energy resolution will
become possible only after the launch of an X-ray microcalorimeter
 on-board of the ASTRO-H mission with
an energy resolution of $\sim$ 4 eV at 6 keV \citep{Mit09}. For example, for the FeXXV line at 6.7 keV, a shift of 
$\sim$ 10 eV emerges during gas motions with a
velocity of 500 km s$^{−-1}$. The tangential component of
gas motions is even more difficult to determine\footnote{\citet{Zhu10} proposed a method for determination 
of tangential gas velocities based on analysis of
the polarization in resonance X-ray lines.}. Another
method for diagnostics of gas motions is based on
analysis of the scattering in bright lines in the spectra
of galaxy clusters (e.g. see the review by \citet{Chu10}). The characteristic amplitudes of gas
motions in galaxy clusters \citep{Chu04,San10} and elliptical galaxies (e.g. see
\citet{Xu02,Wer09}) have been estimated
by comparing the surface brightness profiles in
optically thin and thick lines of the same ion. Finally,
information about the power spectrum of the velocities
of gasmotions can be obtained by considering the
surface brightness or pressure fluctuations. For example,
using a Fourier analysis of the surface brightness
and gas temperature fluctuations in the Coma
cluster (A1656), \citet{Sch04} claimed the
power spectrum of gas motions to be a Kolmogorov
one. A Kolmogorov turbulence power spectrum was
also obtained by \citet{Vog03} using measurements
of the spatial Faraday rotation fluctuations.

In this paper, we are interested in   
what information about the gas motions
we can obtain by considering the distortions of the
surface brightness profiles in resonance lines. We are
interested in the reliability of determining the directions,
amplitudes and spatial scales of gasmotions in
real clusters from resonant scattering observations.

The velocity of ion motions can be represented
as ${\bf V} = {\bf V_{bulk}} + {\bf V_{turb}} + {\bf V_{therm}}$. Here,
  ${\bf V_{bulk}}$ are the
large-scale (bulk) gas motions that affect both the
line broadening and the energy shift at the line center;
${\bf V_{turb}}$ are the gas motions on scales smaller than
any characteristic sizes, in particular, the size of the
region inside which the optical depth in lines $>$ 1 ( microturbulent
motions), that cause only line broadening;
and $\bf V_{\rm therm}$ are the thermal ion motions leading to
the line broadening. Under conditions of galaxy clusters,
the thermal broadening of heavy-element lines can be
much smaller than that for other types of motions.

In this paper, we separately consider the influence
of large- and small-scale gas motions on the line
profiles and resonant scattering. The cluster model
used and details of our calculations are described. The
main results and conclusions are discussed.

\section{LINE PROFILES} 
\subsection{The Influence of Microturbulent Gas Motions on the Line
  Profiles}

As has been mentioned above, we will use the term
``microturbulence'' to describe motions with spatial
scales of velocity variations much smaller than any
characteristic size present in the problem. Such
motions lead to line broadening.

Microturbulence can arise, for example, from the
mergers of clusters, from buoyant bubbles of relativistic
plasma or convection caused by the mixing of thermal
plasma and cosmic rays. Talking about turbulence,
one usually refers to gas motions
with the same velocity dispersion in all directions. However,
the cases when anisotropic turbulence, radial
and tangential, appears are possible. For example,
purely radial motions naturally arise if the energy
from the central active galactic nucleus (AGN) powers shocks and sound waves that propagate through
the intergalactic medium mainly in the radial direction
away from the central source (e.g. see \citet{For05,For07,Fab03,Fab06}). In
contrast, tangential motions can naturally emerge in
stratified atmospheres, where internal waves carry the
energy of vertical motions away from the region of
space under consideration. This can give rise to two dimensional
(tangential) motions \citep{Chu01,Chu02,Reb08}.

Let consider how the anisotropy of turbulent gas motions
changes the spectral line profiles. Let us assume that the gas as a 
whole is at rest at each point of the
cluster and there is a Gaussian ion velocity distribution,
\beq
& \disp P(V_\r,V_{\theta},V_{\phi})=\frac{1}{(2\pi)^{3/2}\sigma_\r
  \sigma_{\theta} \sigma_{\phi}}&\\
&\disp \nonumber\times\exp\left[-\frac{1}{2}\left(\frac{V_\r}{\sigma_\r}\right)^2-\frac{1}{2}\left(\frac{V_{\theta}}{\sigma_{\theta}}\right)^2-\frac{1}{2}\left(\frac{V_{\phi}}{\sigma_{\phi}}\right)^2\right],&
\eeq
where $r$, $\theta$, $\phi$ are three spatial coordinates at a given
point (in the radial and tangential directions) and the
set of three quantities $\Sigma^2=\left (\sigma^2_\r,
\sigma^2_{\theta},  \sigma^2_{\phi}\right)$ characterizes
the velocity dispersion in these directions. Writing
the ion velocity vector in some direction ${\bf m}=(m_\r, m_{\theta},
m_{\phi})$ as ${\bf V}=V{\bf m}$, the probability that the
projection of the ion velocity vector onto the direction $\bf m$ at
a given point of space will be $V$ is
\be
P(V)=\frac{1}{\sqrt{2\pi}\sigma_{\rm
    eff}}\exp\left[-\frac{1}{2}\left(\frac{V}{\sigma_{\rm eff}}\right)^2\right],
\label{eq:distr}
\ee
where $\sigma_{\rm eff}$ can be represented as (given only the
broadening due to turbulence)
\be
\disp
\sigma^2_{\rm eff}=(\sigma_{\rm turb, \rm r} m_{\rm r})^2+(\sigma_{\rm
  turb,\rm \theta} m_{\rm \theta})^2+(\sigma_{\rm turb,\rm \phi} m_{\rm \phi})^2.
\label{eq:sigeff}
\ee
Clearly, for isotropic turbulence, the velocity dispersion
in all directions is the same, i.e.
$\sigma_{\rm turb,\rm r}=\sigma_{\rm turb,\rm \theta}=\sigma_{\rm
  turb,\rm \phi}$. 
For radial turbulence, only the radial
component remains and, accordingly, for tangential
turbulence, the radial component of the velocity dispersion
is zero. Thus, these three cases can be described
as follows:
\be
\Sigma^2=\left\{\begin{array}{rcl}(\sigma_{\rm turb,\rm r}^2,
\sigma_{\rm turb,\rm \theta}^2, \sigma_{\rm turb,\rm \phi}^2), {\rm
  isotropic }\\ \\ (\sigma_{\rm turb,\rm \r}^2, 0, 0), {\rm
  radial}\\ \\ (0, \sigma_{\rm turb,\rm \theta}^2, \sigma_{\rm
  turb,\rm \phi}^2)  , {\rm tangential}\end{array}\right..
\ee
The kinetic energy is related to the velocity dispersion
as  $\disp
\varepsilon_{\rm kin}=\frac{1}{2}\rho(\sigma_{\rm turb,\rm
  r}^2+\sigma_{\rm turb,\rm \theta}^2+\sigma_{\rm turb,\rm \phi}^2)$. 
Fixing the
kinetic energy in all velocity components, we obtain
\be
\Sigma^2=\left\{\begin{array}{rcl}(\frac{2}{3\rho} \varepsilon_{\rm
  kin},\frac{2}{3\rho} \varepsilon_{\rm
  kin},\frac{2}{3\rho}\varepsilon_{\rm kin}), {\rm isotropic
}\\ \\ (\frac{2}{\rho} \varepsilon_{\rm kin}, 0, 0), {\rm
  radial}\\ \\ (0, \frac{1}{\rho} \varepsilon_{\rm  kin},
\frac{1}{\rho} \varepsilon_{\rm kin})  , {\rm tangential}\end{array}\right..
\label{eq:sigE}
\ee
Substituting (5) into (3), we find $\sigma^2_{\rm eff}$ at fixed total
kinetic energy of turbulent motions:
\be
\sigma_{\rm eff}^2=\left\{\begin{array}{rcl}\frac{2}{3\rho}
\varepsilon_{\rm kin}, {\rm isotropic }\\ \\ \frac{2}{\rho}
\varepsilon_{\rm kin}\cos^2(\alpha), {\rm radial}\\ \\ \frac{1}{\rho}
\varepsilon_{\rm kin}\sin^2(\alpha) , {\rm tangential}\end{array}\right.,
\label{eq:sigeffE}
\ee
where $\alpha$ is the angle between the direction of photon
 motion {\bf m} and the radius vector.

Analogously we can derive expressions
for $\sigma_{\rm eff}$ by fixing not the total energy but the velocity
dispersion $\xi$ in a given direction. Assuming
that $\sigma_{\rm turb,\rm r}=\sigma_{\rm turb,\rm \theta}=\sigma_{\rm
  turb,\rm \phi}=\xi$ for isotropic turbulence,
$\sigma_{\rm turb,\rm r}=\xi$, $\sigma_{\rm turb,\rm
  \theta}=\sigma_{\rm turb,\rm \phi}=0$ for radial turbulence,
and $\sigma_{\rm turb,\rm \r}=0$, $\sigma_{\rm turb,\rm
  \theta}=\sigma_{\rm turb,\rm \phi}=\xi$ for tangential
turbulence, we obtain
\be
\sigma_{\rm eff}^2=\left\{\begin{array}{rcl}\xi^2, {\rm isotropic }\\ \\ \xi^2\cos^2(\alpha), {\rm radial}\\ \\ \xi^2\sin^2(\alpha) , {\rm tangential}\end{array}\right..
\label{eq:sigeffV}
\ee

To calculate the spectral line profile from the entire
cluster, let consider a spherically symmetric cluster
and an infinitely thin, homogeneous spherical shell
at distance $r=1$ from the cluster center. The shell
thickness along the line-of-sight is approximately
equal to $\disp\frac{\Delta r r}{\sqrt{r^2-R^2}}=\frac{\Delta r}{\sqrt{1-R^2}}$, where $\Delta r$ is the
shell thickness in the radial direction and $R$ is the
projected radius in the plane of the sky. Let $V$ be
the velocity along the line-of-sight and the Gaussian
ion velocity distribution (2) be valid. We will take
into account only the turbulent line broadening, i.e.,
$\sigma^2_{\rm eff}=(\sigma_{\rm turb,\rm r} m_{\rm r})^2+(\sigma_{\rm
  turb,\rm \theta} m_{\rm \theta})^2+(\sigma_{\rm turb,\rm \phi}
m_{\rm \phi})^2$. At
fixed velocity dispersion in a given direction Eq. (7)
will then be rewritten as
\be
\sigma_{\rm eff}^2=\left\{\begin{array}{rcl}\xi^2, {\rm isotropic }\\ \\ \xi^2(1-R^2), {\rm radial}\\ \\ \xi^2R^2  , {\rm tangential}\end{array}\right..
\label{eq:sigeffpr}
\ee
Substituting (8) into (2) and integrating (2) over the
area $R d R$, we obtain the following line profiles for
isotropic, radial and tangential turbulence:
\beq
&\disp\int\limits_0^1 \frac{\Delta
    r}{\sqrt{2\pi}s}\frac{R}{\sqrt{1-R^2}}&\\
&\nonumber \disp\times\exp\left(\disp-\frac{1}{2}\frac{V^2}{s^2}\right){\rm
    d}R,  {\rm isotropic}& 
\eeq
\beq
& \nonumber\disp\int\limits_0^1 \frac{\Delta
    r}{\sqrt{2\pi}\sqrt{1-R^2}s}\frac{R}{\sqrt{1-R^2}}&\\
&\nonumber\disp\times\exp\left(\disp
-\frac{1}{2}\frac{V^2}{s^2(1-R^2)}\right){\rm d}R,  {\rm radial}&
\eeq
\beq
&\nonumber \disp\int\limits_0^1 \frac{\Delta r}{\sqrt{2\pi}R
  s}\frac{R}{\sqrt{1-R^2}}&\\
&\nonumber\disp\times\exp\left(\disp-\frac{1}{2}\frac{V^2}{s^2 R^2}\right){\rm d}R,  {\rm tangential}&,
\eeq
where $s=\xi$. The expressions for the line profiles at
fixed total kinetic energy are similar but with the substitution
$s=2\varepsilon_{\rm kin}/3\rho, 2\varepsilon_{\rm kin}/\rho,
\varepsilon_{\rm kin}/\rho$ for isotropic, radial and tangential
turbulence, respectively. Figure 1 shows the derived
averaged profiles that agree well with those from
our numerical simulations in the limit of a low
temperature and a large amplitude of gas motions.
Note that the root-mean-square (rms) line widths at
fixed kinetic energy of turbulent motions are identical
for any anisotropy, although the line shapes do not
coincide.

\begin{figure}
\plotone{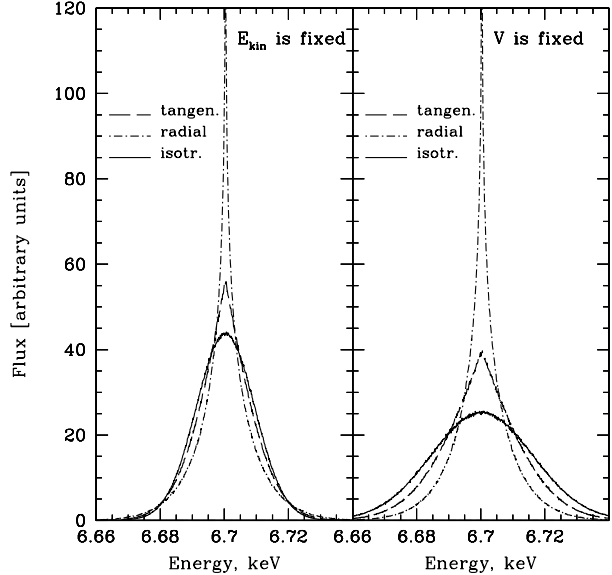}
\caption{Analytical spectral profiles of the helium-like iron line at
  6.7 keV for isotropic microturbulent (solid curves), radial
  (dash–-dotted curves) and tangential (dashed curves) gas motions. 
The line broadening only due to turbulence is taken into account. The 
profiles were calculated at fixed total kinetic energy and fixed 
velocity amplitude of gas motions (9).
}
\end{figure}

The change in line width with distance from the
cluster center for various types of turbulence was considered
by \citet{Reb08}. It was shown that
the linewidth is constant along the radius for isotropic
turbulence, the lines are considerably broader at the
cluster center than at the edges for radial turbulence,
and, on the contrary, the broadest lines are at the
cluster edges for tangential turbulence.

\subsection{The Influence of Large-Scale Gas Motions on the Line
  Profiles}

As was said in the Introduction, large-scale gas
motions cause the lines to be shifted and broadened.
This is easy to see using the results of numerical
simulations of galaxy clusters. Figure 2 shows the
spectral profiles of the helium-like iron line at 6.7
calculated for the model cluster g676, which is an
example of a low-mass, dynamically quiet cold cluster
(see the ``Velocity Field'' Section and Table 1)
by taking into account the large-scale motions and
thermal line broadening. The profiles were calculated
for nine lines-of-sight. For the central panel, the
line-of-sight passes through the cluster center. The
thin solid curves indicate the spectra that emerge in
the case of purely thermal broadening; the profiles
in the presence of gas motions are indicated by the
thick solid curves. The dashed curves indicate the
Gaussian line profile fits. The profiles averaged over
the entire cluster are shown in Fig. 3. We see that
the profiles are broadened and shifted in the
presence of gas motions. Strong deviations from the
Gaussian profile are clearly seen at the cluster periphery,
suggesting the presence of large-scale gas
motions.

Depending on which cluster region we observe
(the cluster center or edge), the line profiles will be
more sensitive to radial or tangential gas motions.
This is demonstrated in Fig. 4, where the spectral
profiles of the helium-like iron line at 6.7 keV are
shown for the model, dynamically quiet cluster g6212
(see the ``Velocity Field'' Section and Table 1) calculated
in three cluster regions: the central region
($\sim$70 kpc in diameter) and two regions at a projected
distance of $\sim$160 kpc from the center ($\sim$145 kpc in
diameter each). We separately consider the influence
of radial (dots) and tangential (dashes) gas motions
and compare with the profiles in the case when the
gas is at rest (solid curves). For this purpose, we
set the radial, tangential, or both velocity components
equal to zero in each cell of the computational volume
and repeated the radiative transfer computation
procedure. We exclude the motion of the cluster as
a whole by subtracting the weighted mean (within a
sphere at the cluster center 50 kpc in radius) velocity.

In the cluster center the scale of gas motions is small compared to the size
of the central region. Therefore, we see an almost
symmetric and slightly broadened line. At the cluster
edges, the spatial scale of gas motions is comparable
to the characteristic size of the region that makes a
dominant contribution to the emission. Therefore,
we see an asymmetric line with a shift of the
central energy. Since the surface brightness and (gas
density) rapidly falls toward the cluster edges, the
regions along the line-of-sight located near the plane
of the sky passing through the cluster center make 
a major contribution to the emission. Obviously, the
observed energy shift for these regions is produced
mainly by tangential motions.

\section{THE INFLUENCE OF THE VELOCITY FIELD ON RESONANT SCATTERING}

The photon scattering probability in a given line is
determined by (a) the optical depth at the line center,
(b) the deviation of the photon energy from the line
energy, and (c) the line width.

\begin{table}
\centering
\caption{ Basic parameters of the galaxy clusters from numerical
simulations.}
\begin{tabular}{@{}rccc@{}}
\hline
Cluster & $M_{\rm vir}$,10$^{14} M_\odot$ & $R_{\rm vir}$, Mpc & $T_{\rm mean}$, keV  \\ \hline
g6212 & 1.61 & 1.43 & 1.5 \\ 
g8 & 32.70 & 3.90 & 13  \\ 
g51 & 19.21 & 3.26 & 8.6  \\ 
g676 & 1.60 & 1.43 & 1.25  \\ 
\hline
\end{tabular}
\end{table}

Turbulent gas motions in the direction of photon
propagation will broaden the line, reducing the optical
depth and the resonant scattering effect (see, e.g.,
\citet{Gil87}). Indeed, the optical depth at
the line center can be found as $\tau=\int\limits_0^{\infty}n_{\i}s_0 d l$, where
$n_i$ is the ion number density and $s_0$ is the scattering
cross section at the line center:
\be
s_0=\frac{\sqrt{\pi}hr_\e cf}{\Delta E_\D},
\ee
Here, $r_{\rm e}$ is the classical electron radius, $f$ is the
oscillator strength of a given transition, and $\Delta E_{\rm D}$ is
the Doppler line width. The line broadening $\Delta E_{\rm D}$ can
be divided into two components: the purely thermal broadening and the
broadening due to turbulent
motions, i.e.,
\be
\disp \Delta E_{\rm D}=\disp \frac{E_0}{c}\sqrt{2(\sigma^2_{\rm
    therm}+\sigma^2_{\rm turb})}.
\label{eq:deltaE}
\ee
Here, $\sigma_{\rm turb}$ is the velocity dispersion due to turbulent
motions and $\sigma_{\rm therm}$ is the thermal dispersion of ion velocities:
\be
\disp \sigma^2_{\rm therm}=\frac{kT}{Am_{\rm p}},  
\label{eq:sturb}
\ee
$E_0$ is the energy at the line center, $k$ is the
Boltzmann constant, $m_{\rm p}$ is the proton mass, $A$ is the
atomic weight of the element, and $c$ is the speed of
light. For heavy elements $A$ is large (e.g., $A = 56$
for iron), therefore the thermal broadening of
heavy-element lines is noticeably suppressed, while the broadening through gas motions is the same for
all lines.

\begin{figure}
\plotone{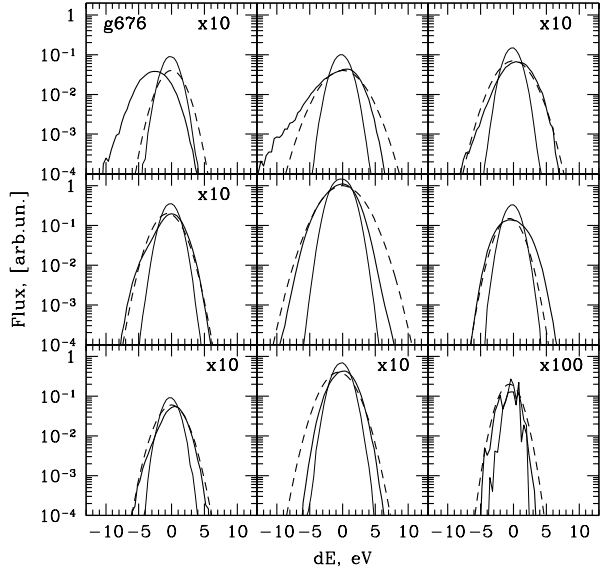}
\caption{Profiles of the helium-like iron FeXXV line at 6.7 keV for nine lines-of-sight in the model cluster g676 (see the table).
The cluster was divided into nine identical parts; the size of the entire cube is $1\times1\times1$ Mpc. For the central panel, the line
of sight passes through the cluster center. The thick, thin and dashed curves indicate, respectively, the line profiles shifted
and broadened through gas motions, the profiles emerging only in the case of thermal line broadening, and the Gaussian line
profiles broadened through the mean temperature and $V_{rms}$ in each region.
}
\end{figure}

\begin{figure}
\plotone{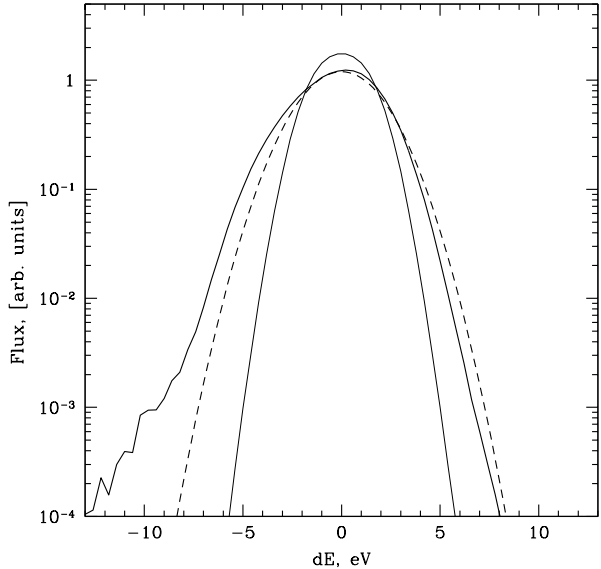}
\caption{Profiles of the helium-like iron FeXXV line at 6.7 keV
  averaged over the entire model cluster g676. The notation is the same as that in Fig. 2.
}
\end{figure}
The presence of large-scale gas motions leads to
a change in the resonant scattering cross section.
In the reference frame of the gas, the cross section
is proportional to $\displaystyle {\rm exp}\left \{-\frac{\left(E\left [1-\frac{\bf{({\bf V} {\bf m})}}{c}\right ]-E_0\right )^2}{2\sigma ^2}\right\}$,
where $E$ is the photon energy in the reference frame
of the cluster, $E_0$ is the transition energy, ${\bf V}$ is the
velocity of large-scale gas motions, ${\bf m}$ is the photon
propagation direction, and $\sigma$ is the Gaussian line
width determined by the thermal broadening and the
broadening through turbulent gas motions.

\section{CLUSTER MODEL AND SCATTERING CALCULATION}

At present, the gas density and temperature distributions
in nearby X--ray bright galaxy clusters, such as Perseus cluster, are well known.
At the same time, there is almost no information
about the properties of the gas velocity field. It seems
natural to supplement the density and temperature
measurements with the results of numerical hydrodynamic
velocity calculations and to use such a combined
model to model the resonant scattering.

\subsection{Gas Temperature and Density Distributions}

The resonant scattering calculations were performed
for the Perseus galaxy cluster A426, whose
electron temperature and density profiles were taken
from \citet{Chu03}. Correcting the profile for
the Hubble constant $H_0 = 72$ km s$^{-−1}$ Mpc$^1$, we find
the electron density in cm$^{-−3}$ as
\be
n_\e=\frac{4.68\cdot10^{-2}}{\left(1+\left(\frac{r}{56}\right)^2\right)^{3/2\cdot1.2}}+\frac{4.86\cdot10^{-3}}{\left(1+\left(\frac{r}{194}\right)^2\right)^{3/2\cdot0.58}}.
\ee
The temperature distribution in keV is
\be
T_\e=7\frac{1+\left(\frac{r}{100}\right)^3}{2.3+\left(\frac{r}{100}\right)^3},
\ee
where $r$ is in kpc. The iron abundance is assumed
to be constant in the  cluster and equal to 0.5
of the solar one from the tables by \citet{And89} and 0.74 of the solar if one uses the
newer tables by \citet{Asp09}.

The mean temperature in the Perseus cluster is
about 5–-6 keV. At such temperatures, the strongest
line in the spectrum is the $K_\alpha$ line of helium-like iron
at 6.7 keV that corresponds to the $1s^2(^1S_0)–-1s2p(^1P_1)$
transition with an oscillator strength $\sim$ 0.7. The
optical depth in this line from the cluster center to
infinity is $\sim$ 3.

\subsection{Velocity Field}

The energy resolution of modern X-ray telescopes
does not allow the gas velocity field in galaxy clusters
to be measured directly. To make the assumption
about the velocity field, we will use the results of numerical
cluster simulations taken from the cosmological
calculations of the large-scale structure \citep{Dol05,Spr01}. The numerical
simulations give the three-dimensional structure of
the gas density and temperature and three velocity
components determined in a cube with a size of several
Mpc. We considered nine clusters differing in
mass. The rms velocity amplitudes in the coldest
and hottest clusters are $\sim$ 200 and $\sim$ 1000 km s$^{-−1}$,
respectively.

\begin{figure}
\plotone{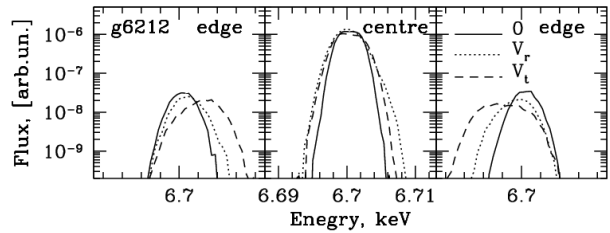}
\caption{Spectral profile of the helium-like iron FeXXV line at 6.7 keV calculated for the model cluster g6212 (see the table) at
the cluster center ( the size is $\sim$70 kpc) and in two regions at a distance of 160 kpc from the center (the size is $\sim$145 kpc). The
solid, dotted, and dashed curves indicate, respectively, the profiles when the gas is at rest, in the case of radial gas motions, and
in the case of tangential gas motions.
}
\end{figure}

\begin{figure}
\plotone{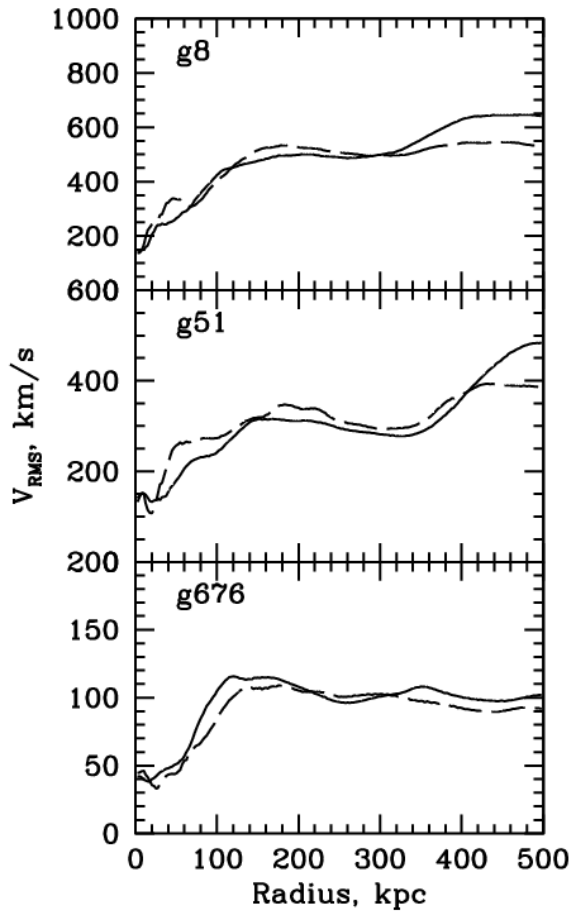}
\caption{Rms amplitudes of the radial velocity of gas motions (dashed
  curves) and rms amplitudes of the tangential velocity $\sqrt{(V_\theta^2+V_\phi^2)/2}$ (solid curves) versus distance from
  the cluster center for three model clusters taken from numerical simulations.
}
\end{figure}
\begin{figure}
\plotone{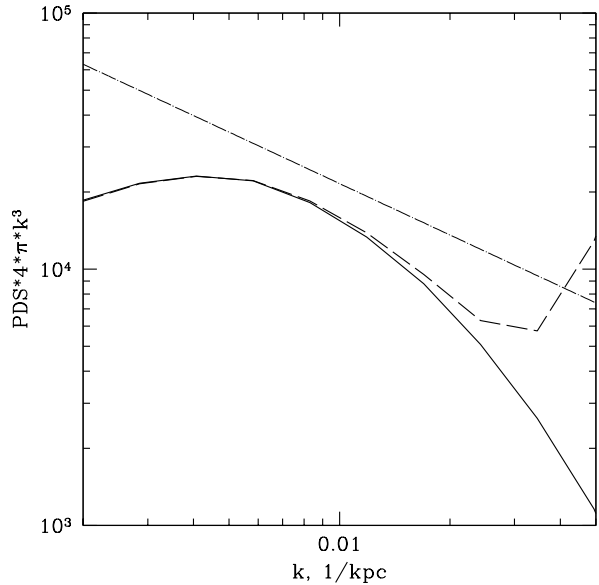}
\caption{Power spectrum of the velocities of gas motions
in the cluster g676 (see the table) from numerical
simulations. To construct the spectra, we used the results
of SPH simulations smoothed adaptively (solid curve)
and non-adaptively (dashed curve). The
Kolmogorov power spectrum $\sim k^{-−2/3}$ is indicated by the
dash–-dotted curve.
}
\end{figure}

To make the conclusion about the directions of
motions, let us compare the rms amplitude of the
radial velocity $V_{\rm r}$ with that of the tangential one
$\sqrt{(V_\theta^2+V_\phi^2)/2}$.
 We exclude the motion of the cluster
as a whole by subtracting the weighted mean (within
a sphere at the cluster center 50 kpc in radius) velocity. 
The rms amplitude $V_{\rm rms}$ was calculated in
spherical shells with radius $r$ and thickness $\Delta r=0.1 r$ 
separately for the velocity components along
and perpendicular to the radius. The results for the
available nine clusters show that the gas motions are
almost isotropic. Figure 5 presents the results for
three clusters. The rms amplitudes of the velocities
of radial gas motions are indicated by the dashed
curves; the solid curves indicate the amplitudes for
each velocity component in the tangential direction.
We see that the curves differ insignificantly at radii
$< 300 - 400$ kpc and, hence, the amplitudes of the
gas motions are identical in all directions.

Although the assumption that the gas motions
are isotropic is justified by our numerical simulations,
it should be remembered that in reality the gas
motions are not necessarily isotropic. For example,
considering the velocities of gas filaments in H$_\alpha$ in the
Perseus cluster, \citet{Hat06} showed that the
gas at distances larger than 30 kpc from the central 
galaxy (i.e., where the resonant scattering is significant)
entrained by AGN-inflated plasma bubbles is
involved predominantly into radial motions.

To make reasonable assumptions about the properties
of the velocity field in our radiative transfer calculations,
we analyzed the power spectra found
in hydrodynamic simulations of cluster formation. An
example for one cluster is shown in Fig. 6. Details
of the method for calculating the power spectrum are
given in \citet{Are11}. Most of the numerical
simulations suffer from an insufficient dynamic range
of wave numbers needed to reliably determine the
shape of the power spectrum. In particular, adaptive
smoothing, which is especially significant for the
SPH method (the solid curve in Fig. 6), leads to a
cutoff in the power spectrum at large $k$. At the same
time, non-adaptive smoothing leads to
a sharp rise of the power spectrum at large wave numbers,
which is related to Poissonian noise on these
scales, in particular, when there are
no particles at all within the cell. The true power
spectrum at large $k$ must lie between the dashed and
solid curves shown in Fig. 6. The cutoff in the power
spectrum at small wave numbers is related, in particular,
to the choice of the size of the computational
volume centered on the cluster. We discuss this effect
in a separate publication \citep{Zhu11}.
In Fig. 6, the dash–-dotted curve indicates the Kolmogorov
power spectrum\footnote{The Kolmogorov spectrum has such a slope when the total
energy associated with the motions with a given scale of
wave numbers $k$ is considered.} $\sim k^{-−2/3}$. One can see that the
assumption about a Kolmogorv spectrum appears as
a reasonable compromise, although the Kolmogorov
slope in the simulated clusters is observed in a fairly
narrow range of wave numbers.

Yet another problem can be associated with an
insufficient resolution of SPH simulations on small
scales. The scales that are resolved in simulations
range from tens of kpc at the cluster center to several
hundred kpc at the edges. An additional spatial scale
arises in the scattering problem —- the characteristic
photon mean free path in scattering, i.e., the size at
which an optical depth of the order of unity is accumulated.
It is useful to compare this scale with the
resolution of SPH simulations. For the cluster g676,
the power spectrum is shown in Fig. 6; we see that on 
$k \le 0.02$ kpc$^{-−1}$ the simulations are resolved. On
shorter scales (large wave numbers), the calculations
of the velocity field with an adaptive window (solid
curve) and a non-adaptive window (dashed curve)
give large difference in results. This discrepancy means
that the simulation resolution limit was reached. The
corresponding spatial scale\footnote{We use the
  relation $k = 1/l$ between the spatial scales and wave numbers without the factor $2\pi$.} is $\sim$ 50 kpc. In this case,
an optical depth of the order of unity near the cluster
center is accumulated at a size  $\displaystyle 1/(n_i s_0)\sim 150 $ kpc.
Thus, the resolution of our simulations is sufficient for
our purposes.

In the subsequent simulations, we assumed the
power spectrum of the velocity fluctuations to be
$k^3P(k) \sim k^{−-2/3}$.

\subsection{Monte Carlo Simulations of Scattering}

To simulate multiple scattering, we used the
Monte Carlo method \citep{Poz83}.
Details of the simulations of scattering in lines
are discussed in \citet{Saz02,Chu04} and \citet{Zhu10}. The line
energies and oscillator strengths were taken from the
ATOMDB\footnote{http://cxc.harvard.edu/atomdb/WebGUIDE/index.html.} and NIST Atomic Spectra Database\footnote{http://physics.nist.gov/PhysRefData/ASD/index.html.}
databases.

During the scattering, the photon direction $\bf m$
is selected by taking into account the scattering
phase function, which is a combination of Rayleigh and
isotropic scattering phase functions \citep{Ham47,Cha50}. For isotropic scattering, the
new photon direction $\bf m'$ is drawn randomly. For Rayleigh scattering, the probability that the photon
after its scattering will propagate in the direction $\bf m'$ is $P(\bf m' ,
\bf e')\propto(\bf e' \cdot \bf e)^2$, where the direction of the
electric field is $\bf e'=(\bf e - \bf m'
  \cos(\alpha))/(\sqrt{1-\cos^2(\alpha)}) $
and $\alpha$ is the angle between the electric vector $\bf e$ before
the scattering and the new photon direction $\bf m'$, i.e.
$\cos(\alpha)=(\bf e \cdot \bf m')$.

The initial photon position (or initial weight)
is chosen in accordance with the volume emissivity
of various cluster regions. To calculate the
line emissivity, we used the APEC code \citep{Smi01}. The ionization balance was taken from
\citet{Maz98}. For a randomly chosen
initial direction of photon propagation $\bf m$, we find the
photon energy $E$ with a Gaussian distribution with
mean $E_0$ and standard deviation $\Delta E_\D/\sqrt{2}$, where
$\Delta E_\D$ was found from Eq. (11). Since the gas density
is low, the pressure effects on the line broadening
are neglected. We also neglect the contribution
from the radiative decay of levels to the broadening
by assuming the levels to be infinitely thin. When
simulating the scattering process, we find the velocity
of the scattering ion in such a way that the photon
energy in the reference frame of the ion is exactly
equal $E_0$ and the scattering occurs. Thus, the ion
velocity in the direction of photon motion is 
$V_{\rm ion}=\left(1-\frac{E_0}{E}\right) c$. 
We find the other two ion velocity
components $V_{{\rm ion} 1}$ and $V_{{\rm ion} 2}$ in the directions $\bf m_1$
and $\bf m_2$ orthogonal to $\bf m$ as $(\bf{{\bf V}_{\rm gas}},
\bf{{\bf m}_1})$$+ V_{\rm gauss}$ and
$(\bf {{\bf V}_{\rm gas}, {\bf m}_2})$$+ V_{\rm gauss}$, where $V_{\rm
  gauss}$ is the velocity with
a Gaussian distribution. We take into account the
velocity dispersion in the directions $\bf m_1$ and $\bf m_2$ according
to Eqs. (7) or (6). As a result, the velocity of the
scattering ion is ${ \bf V}_{\rm tot}=V_{\rm ion} {\bf m} + V_{{\rm
    ion} 1} {\bf m_1} + V_{{\rm ion} 2} {\bf m_2}$.
Accordingly, after the selection of a new photon
direction by taking into account the scattering phase
matrix, we find the photon energy after the scattering
as $E=E_0 \left(1+({\bf V}_{\rm tot}\cdot{\bf m'}\right)/c)$. In this case, we
neglect the change in photon direction when going
from the laboratory reference frame to the reference
frame of the ion and vice versa, because $V/c \ll 1$.
This approximation also assumes the velocity field and
the gas distribution to be constant on the time scales
of the photon propagation through the cluster.

\begin{figure*}
{\centering \leavevmode
\epsfxsize=1.\columnwidth \epsfbox[70 190 620 680]{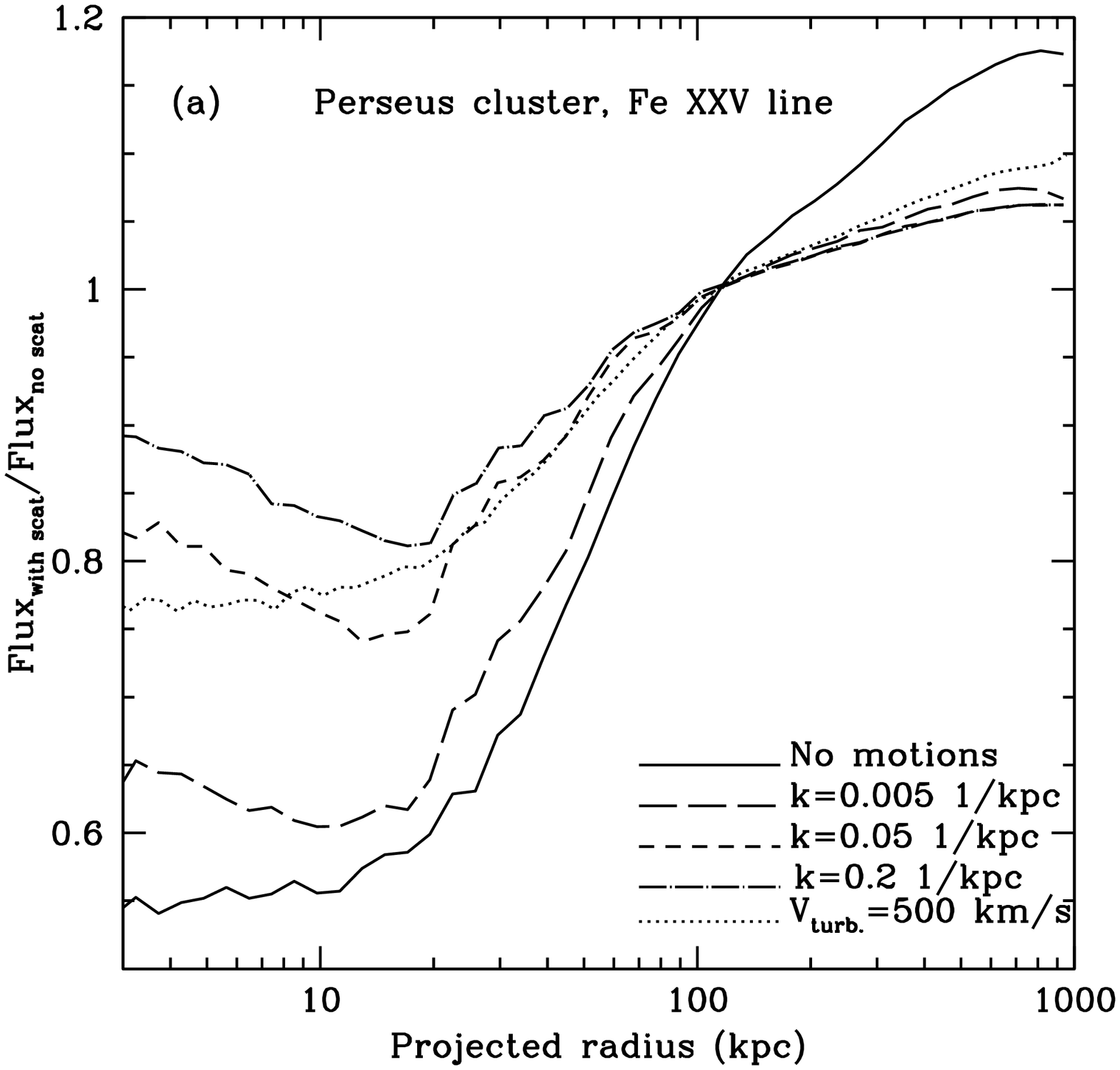}
\epsfxsize=1.\columnwidth \epsfbox[70 190 620 680]{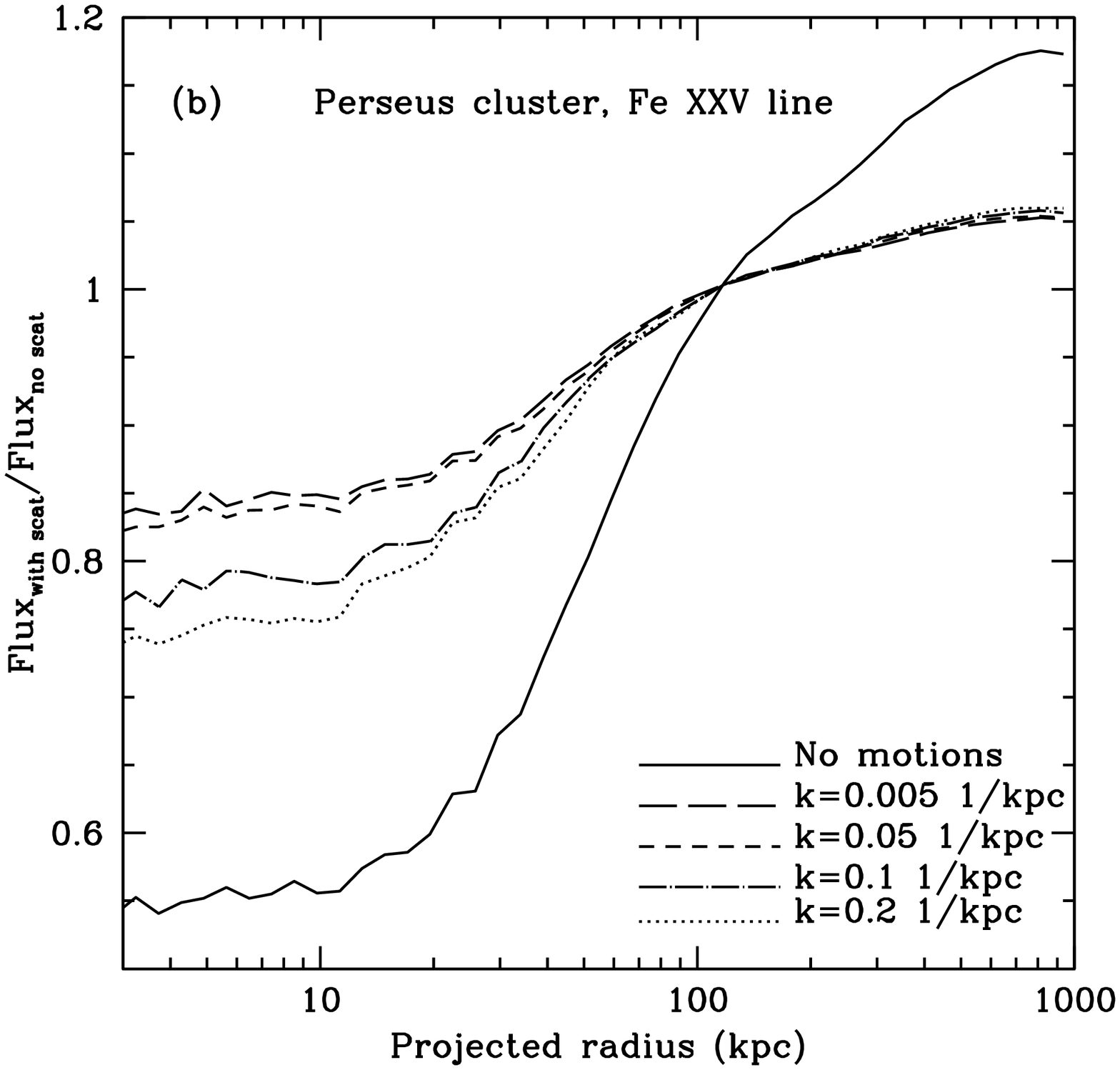}\hfil
\epsfxsize=0.95\columnwidth \epsfbox[100 190 620 730]{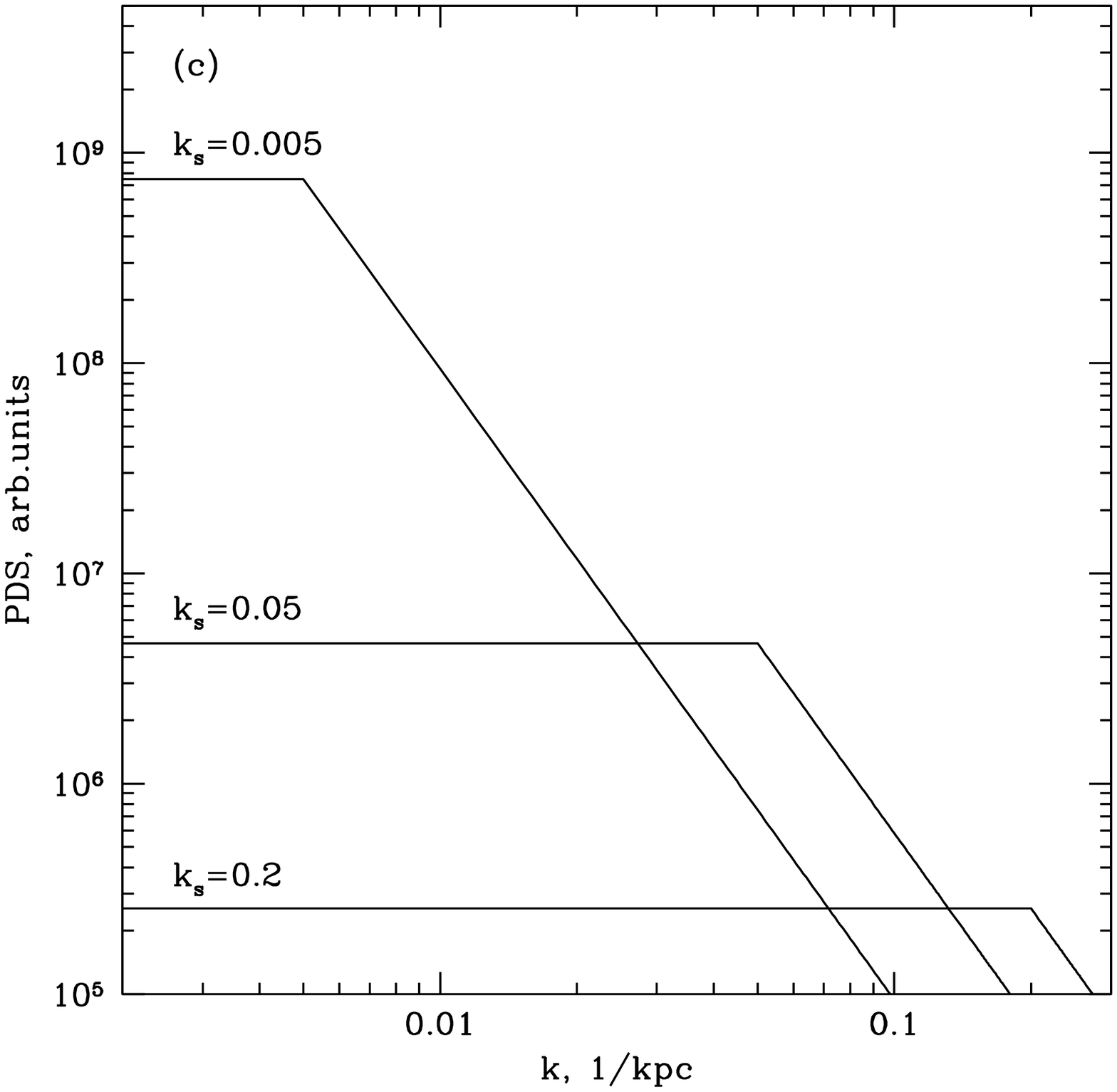}
\epsfxsize=0.95\columnwidth \epsfbox[70 190 590
  730]{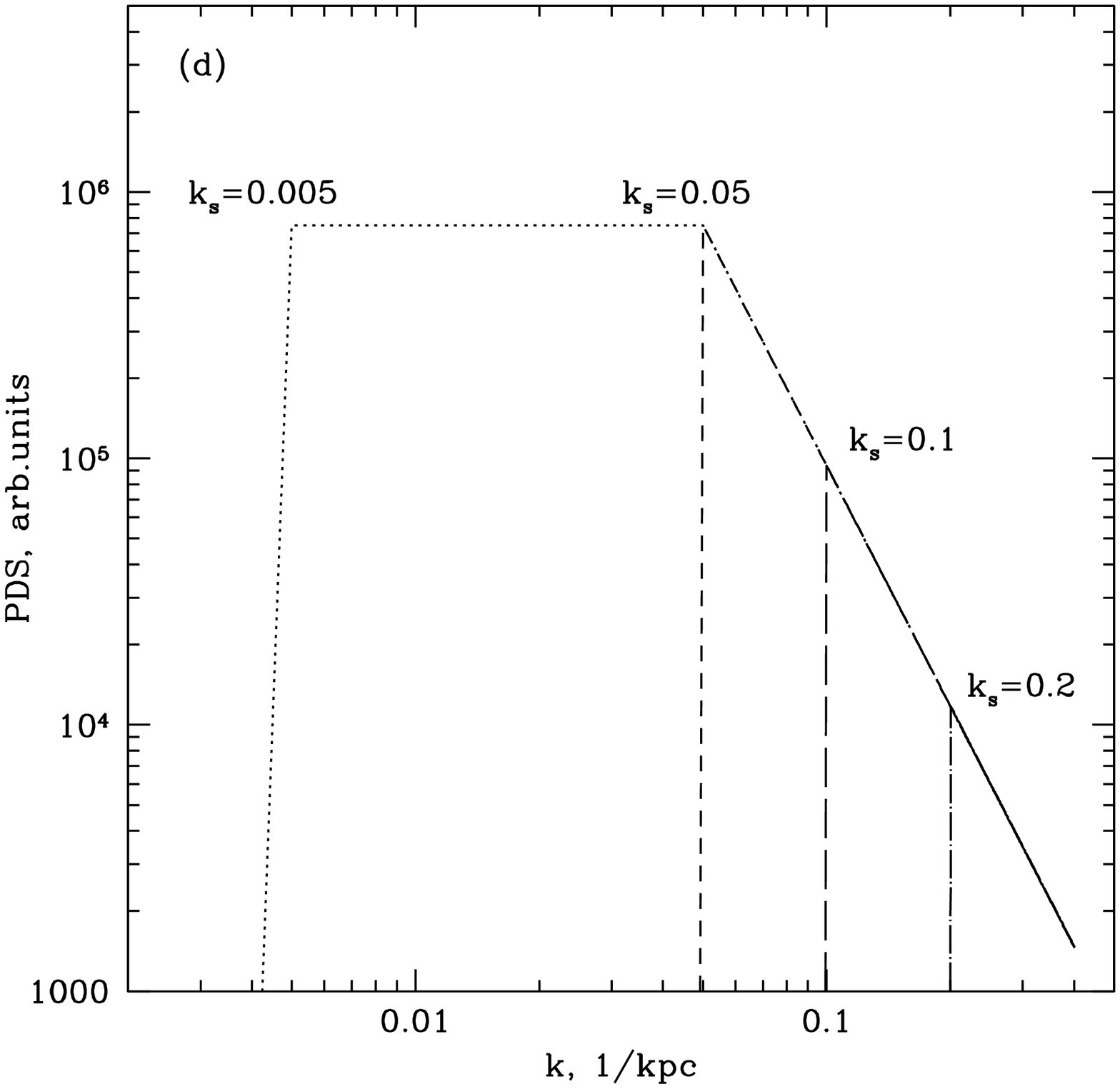}}
\vspace{0.5cm}
\caption{The upper panels: the ratio of the fluxes in the helium-like iron FeXXV line at 6.7 keV with and without scattering in
the cluster A426 for gas motions on various scales. The left upper panel: the sensitivity to large-scale motions (case B, see the
``Results'' Section). The curves correspond to different values of
k$_{\rm s}$ (see the lower left panel). The curve for microturbulence
is presented for comparison. The right upper panel: the sensitivity to small-scale motions when the power spectrum (see the
lower right panel) of the velocities of gas motions is cut off at various k (case C, see the ``Results'' Section). The lower panels:
the power spectra of the velocities of gas motions in galaxy clusters assumed in the calculations. The left lower panel: the
power spectrum is flat at small $k_{\rm s}$ (=0.005, 0.05, and 0.2) and then falls off as the Kolmogorov one ( see the ``Results'' Section).
The right lower panel: the power spectrum for $k_{\rm s}$ = 0.05 (see the left panel) is cut off at $k >$ 0.005, 0.05, 0.1 and 0.2 (see the
``Results'' Section).
}
\end{figure*}

\section{RESULTS}

To calculate the influence of the velocity field on the
scattering, we generate a random realization of the
power spectrum with a given shape and normalization
for each velocity component and make the inverse
Fourier transform to obtain the velocity field with
given properties.

We performed our calculations for three velocity
fields.

A. {\bf The regime of microturbulence:} there are
no large-scale motions, microturbulence leads to line
broadening through isotropic, radial, and tangential
turbulence with $V_{rms} = 500$ km s$^{-−1}$.

B. {\bf The sensitivity to large-scale gas motions:}
the power spectrum is flat at small $k$ up to $k_{\rm s}$ and
then falls off as the Kolmogorov one (see Fig. 8c). In
this case, the flat power spectrum at small $k$ implies
the absence of large-scale motions with sizes $> 1/k_{\rm s}$.
The bulk of the power is related to wave numbers
of the order of $k_{\rm s}$. Our calculations were performed
for several values of $k_{\rm s}$ ($l_{\rm s}$): 0.005 kpc$^{-−1}$ (200 kpc),
0.05 kpc$^{−-1}$ (20 kpc), and 0.2 kpc$^{-−1}$ (5 kpc). The
rms amplitude of the velocity $V_{rms}$ is 500 km s$^{-−1}$ in
all cases.

C. {\bf The sensitivity to small-scale motions:} the
power spectrum is cut off at $k > k_{\rm s}$ (see Fig. 8d); in
this case, the decrease in the characteristic amplitude
of the velocity is compensated by the line broadening,
i.e., the power is transferred to microturbulence
(the shortest scales).

In all three cases, the total energy in large- and
small-scale motions is the same. The results are
presented in Figs. 7 and 8 (corresponding to cases A,
B, and C).

Figure 7 shows the ratio of the fluxes in the
helium-like iron line at 6.7 keV with and without scattering in the case when there are no gas motions
and in the case of different microturbulence
designated in the figure (case A). The case of fixed
velocity amplitude is indicated by the thick curves.
The corresponding results of our calculations at fixed
total energy are indicated by the thin curves. We see that when the gas is at rest, the optical depth
is maximal, the line flux at the cluster center
is strongly suppressed and the flux at the edges increases
due to the scattering. When the gas motions
are tangential, it is clear that the optical depth in the
line calculated in the radial direction will not change. 
Since the photons produced in the cluster center  moving in the radial direction make
a major contribution to the scattering, the changes
in brightness profiles are insignificant. The radial gas
motions that directly affect the optical depth for the
photons emerging from the cluster center and that
reduce considerably the scattering efficiency have the
strongest effect on the resonant scattering. Note that
the optical depth in this case decreases by almost a
factor of 3. Clearly, the case of isotropic turbulence is
intermediate.

Considering large-scale gas motions (case B),
note that when the bulk of the power of the motions
is on large scales, the scattering is almost as efficient
as that in the case when the gas is at rest (Fig. 8a,
the curve with long dashes). In this case, a bulk
motion of large gas volumes (the scales of the motions
are larger than $r_{\rm c}$) inside which the scattering
 actually takes place. An increase in $k_{\rm s}$, i.e.
assuming the power spectrum to be flat at small $k$
(with the total energy of the motions conserved), leads
to suppression of the role of large-scale motions and
to an increase in the velocity dispersion on small scales.
This immediately leads to a decrease in the optical
depth in the line and resonance scattering effect is less strong (the curve with short dashes
and the dash–-dotted curve; the dotted curve is drawn
for comparison with the case of isotropic microturbulence).

We also see from Fig. 8b that the earlier we ``cut
off'' the power spectrum (case C), i.e. the larger the 
power is in the small-scale motions affecting the line
broadening, the weaker is the resonant scattering.

\section{CONCLUSIONS}

\begin{figure}
\plotone{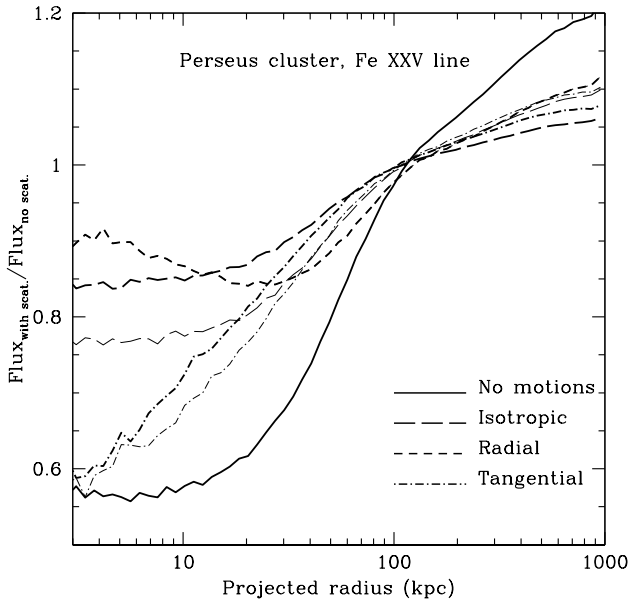}
\caption{Ratio of the fluxes in the helium-like iron FeXXV
line at 6.7 keV with and without scattering in the cluster
A426 for isotropic, radial, and tangential microturbulent
gas motions (case A). The thick and thin curves
correspond to the cases of fixed velocity amplitude and
fixed total kinetic energy, respectively.
}
\end{figure}

\begin{figure}
\plotone{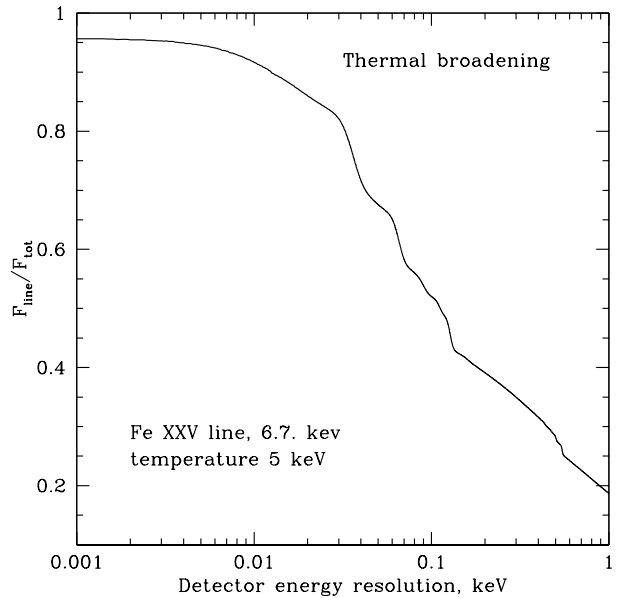}
\caption{Ratio of the FeXXV line flux to the total flux, including the continuum and neighboring lines, versus detector energy
resolution for a cluster with a mean temperature of 5 keV. Only the thermal line broadening is taken into account.
}
\end{figure}

We considered the influence of large- and small-scale
anisotropic gas motions on the resonant scattering
in the FeXXV line at 6.7 keV for the Perseus cluster
 as an example. The model velocity field
is taken from the results of
hydrodynamic simulations of galaxy cluster formation
(\citet{Dol05}).

We showed that (1) the resonant scattering is
sensitive mainly to small-scale gas motions, (2) it is
particularly sensitive to radial motions, (3) large-scale
gas motions affect mainly the shift of the line center
and (4) the directions of small-scale motions can be
estimated by considering the broadening of spectral
lines and the resonant scattering in lines simultaneously.

The sensitivity to anisotropy of microturbulent
motions is illustrated by the following example: at
fixed total kinetic energy corresponding to $V_{\rm rms} =
500$ km s$^{-−1}$, the expected decrease in FeXXV line
flux at the center ($R < 10$ kpc) of the cluster A426 is
a factor of 0.77, 0.9 and 0.61 for isotropic, radial,
and tangential turbulence, respectively. At fixed
velocity dispersion in one direction (Eq. (7)), the
corresponding decreases are a factor of 0.84, 0.9, and
0.63, respectively. In other words, if a significant
anisotropy of small-scale gas motions is allowed,
then the same ratio of the line fluxes corresponds to 
different characteristic velocity amplitudes.
For example, for the case considered above (Fig. 7),
approximately the same ratio of the line fluxes from a
region 10 kpc in radius arises at $V_{\rm rms} = 500$ km s$^{−-1}$
for isotropic motions, $V_{\rm rms} \approx 200$ km s$^{−-1}$ for radial
motions, and $V_{\rm rms} \approx 1700$ km s$^{−-1}$ for tangential
motions. Note that such a large difference is related,
in particular, to a nonlinear dependence of the line
flux ratio on the velocity of gas motions. For a
larger region ($r <$ 30 kpc), a comparable line flux
ratio arises at $V_{\rm rms} = 500$ km s$^{−-1}$ (isotropic motions),
$V_{\rm rms} \approx 300$ km s$^{−-1}$ (radial motions), and 
$V_{\rm rms} \approx 1200$ km s$^{-−1}$ (tangential motions). In this case, the
difference between the velocity amplitudes is about a
factor of 4. Of course, the presence of purely radial or purely
tangential motions in clusters is unlikely and the typical uncertainty is not so
great. Note also that the results of our calculations,
obtained under the assumption of isotropic gas motions, can
be used to set conservative upper limits on the
amplitude of purely radial motions, while robust
constraints on the amplitude of tangential motions
are difficult to obtain from resonant scattering observations.

Large-scale gas motions affect weakly the scattering
efficiency. For example, we see from Fig. 8a that
if the whole power of the motions is on scales of
200 kpc, then the scattering efficiency at the cluster
center is a factor of 1.13 lower than that in the case
when the gas is at rest. In contrast, if the bulk of the
power is concentrated on small scales, for example,
on scales of $\sim$ 5 kpc, then the scattering efficiency
decreases by a factor of 1.56 (see Figs. 8a and 8b).

Note that the considered line of helium-like iron at 6.7 keV
has bright forbidden and intercombination
lines as well as satellites. The optical depths of
the forbidden and intercombination ($1s^2 −- 1s2p(^3P_0)$)
lines are almost zero, while the oscillator strength of
the intercombination ($1s^2 −- 1s2p(^3P_1)$) line is a factor
of 10 smaller than that of the resonance line and its
optical depth is $\sim$ 0.26. The satellites correspond to
the transitions from excited states and have a negligible
optical depth at a low matter density. Thus,
only the resonance transition is essentially involved in
the scattering itself. Clearly, the lower the detector
energy resolution, the more difficult to measure the
suppression of the flux in the resonance line through
scattering. In Fig. 9, the ratio of the 6.7 keV line flux
to the total flux, including the continuum and neighboring
lines, $F_{\rm line}/F_{\rm tot}$ is plotted against the detector
energy resolution. We see that the 6.7 keV line flux
for a cluster with a mean temperature of $\sim$ 5 keV
accounts for 50\% of the total incoming flux in the
energy range $E_0 \pm \Delta/2$ even at an energy resolution
$\Delta \sim 100$ eV.

Apart from their influence on the surface brightness
profile, gas motions change the degree of polarization
in X-ray lines, which results from resonant
scattering in the presence of a quadrupole moment in
the radiation field. \citet{Zhu10} showed
that in the presence of isotropic microturbulent or
large-scale gas motions, the
degree of polarization in galaxy clusters can decrease
by several times. Clearly, depending on the directions
of microturbulent motions, the degree of polarization
will also change.

\section{Acknowledgements} 
This study was supported by the RAS Programs
P19 and DPS 16, the Program for Support of Leading
Scientific Schools (grant NSh-5069.2010.2), and
the Russian Foundation for Basic Research (project
no. 09-02-00867-a). S. Sazonov is grateful to
the Dynasty Foundation for support. I. Zhuravleva
is grateful to the International Max Planck Research
School (IMPRS).

\label{lastpage}
\end{document}